\documentclass[]{elsarticle}
\usepackage{amsmath,amssymb}
\usepackage{bm}
\usepackage{graphicx}
\usepackage{hyperref}
\usepackage{color}

\hypersetup{
pdfauthor = {Daniel, Mucio, Christopher and Claudine},
pdftitle = {\texorpdfstring{$s$}{s}- and \texorpdfstring{$d$}{d}-wave superconductivity in a two-band model \today.},
}
\begin{document}
\title{\texorpdfstring{$s$}{s}- and \texorpdfstring{$d$}{d}-wave superconductivity in a two-band model}

\author[IME]{Daniel Reyes\corref{cor1}}
\ead{daniel@cbpf.br}
\author[CBPF]{Mucio A. Continentino}
\author[IFUFRGS]{Christopher Thomas}
\author[IN,CNRS]{Claudine Lacroix}

\cortext[cor1]{Corresponding author}

\address[IME]{Instituto Militar de Engenharia - Pra\c{c}a General Tib\'{u}rcio, 80, 22290-270, Praia Vermelha, Rio de Janeiro, Brazil}
\address[CBPF]{Centro Brasileiro de Pesquisas F\'{\i}sicas - Rua Dr. Xavier Sigaud, 150-Urca, 22290-180,RJ-Brazil}
\address[IFUFRGS]{Instituto de F\'{i}sica, Universidade Federal do Rio Grande do Sul, C.P. 15051, 91501-970 Porto Alegre-RS, Brazil}
\address[IN]{Univ. Grenoble Alpes, Institut N\'eel, F-38000 Grenoble, France}
\address[CNRS]{CNRS, Institut N\'eel, F-38000 Grenoble, France}

\date{\today}
\begin{abstract}
Superconductivity in strongly correlated systems is a remarkable phenomenon that attracts a huge interest.  The study of this problem is relevant for materials as the  high $T_c$ oxides,  pnictides and heavy fermions. In this work we study a realistic model that includes the relevant physics of superconductivity in the presence of strong Coulomb correlations. We consider a two-band model, since most of these correlated systems have electrons from at least two different atomic orbitals coexisting at their Fermi surface.  The Coulomb repulsion is taken into account through a local repulsive interaction. Pairing is considered among quasi-particles in neighbouring sites and we allow for  different symmetries of the order parameter. In order to deal with the strong local correlations, we use the well known slave boson approach that has  proved very successful for this problem. Here we are interested in obtaining the zero temperature properties of the model, specifically its phase diagram and the existence and nature of superconducting quantum critical points. We show that these can arise by increasing the mixing between the two bands. Since this can be controlled by external pressure or doping, our results have a direct relation with experiments. We show that the superconductor-to-normal transition can be either to a metal, a correlated metal or to an insulator. Also we compare the relative stability of $s$ and $d$-wave paired states for different regions of parameter space and investigate the BCS-BEC crossover in the two-band lattice model as function of the strength of the pairing interaction.
\end{abstract}


\maketitle

\section{Introduction}\label{secint}

The study of superconductivity in strongly correlated electron systems (SCES) is one of the most exciting area in condensed matter physics. It encompasses many interesting materials and also those which are most promising for practical applications.
The high $T_c$ cuprates, the iron pnictides and the heavy fermion materials are among the superconductors where the electronic correlations play an important role. The majority of these systems are multi-band with electrons from different atomic orbitals coexisting at their common Fermi surface. A model to describe these materials must include this feature and also the strong local Coulomb repulsion among electrons of the narrow $d$ or $f$-bands. The most favorable conditions for the appearance of superconductivity involve an attractive interaction which pairs quasi-particles in neighboring sites and in this way avoids the strong on-site repulsion. This type of pairing is often attributed to intersite exchange interaction. It allows for different symmetries of the superconducting order parameter and experimentally it is well known that in the case of the cuprates, they adopt a $d$-wave pairing in which on-site pairing vanishes.

An interesting and challenging observation in strongly correlated superconductors is the existence of a new type of quantum phase transition associated with a superconducting quantum critical point (SQCP)~\cite{Sheila,greene}. Varying an external parameter, such as, doping or pressure these systems can be driven to a non-superconducting state, in most of the cases a metallic state.
This metallic state close to the SQCP is susceptible to the fluctuations arising from the proximity of the superconducting state and these may give rise to non-Fermi liquid type of behavior~\cite{Sheila}. It is a challenging new problem, both experimental and theoretical, to characterize the universality classes of these new SQCP and compare them with the well studied case of magnetic quantum critical points.

 In multi-band systems, hybridization, which is due to the overlap of wave functions, is sensitive to doping or external pressure. In this way it can be modified and to act as an important control parameter, which allows exploring the phase diagram of different materials. However, a complete description of a given system requires including additional ingredients in the model.
For example, the k-dependence of hybridization is important: it is known both, experimentally~\cite{Sheila,Bauer}, and theoretically~\cite{Igor}, that when hybridization is constant or has even-parity in k-space, it acts in detriment of superconductivity and can even destroy it at a SQCP. On the other hand, anti-symmetric or odd-parity hybridization, which occurs when hybridization mixes orbitals with different parities, enhances superconductivity~\cite{Fernanda,Daniel15}. It turns out that this effect is important since it includes the cases of s-p, p-d and d-f orbitals mixing, which are relevant for semiconductors, pnictides, oxide superconductors and heavy fermion materials. We found similar behavior for our model for s and d-wave superconductors depending on the band filling and the intensity of the hybridization~\cite{Daniel16}.

The Anderson lattice model (ALM) is generally accepted as an appropriate model for describing  both magnetic and superconducting instabilities~\cite{Werner,Sacramento,Dorin,Perez,Uinfty} in strongly correlated multi-band systems. Many studies have shown that the pairing between the nearly localized $d$ of $f$-electrons is the responsible for the appearance of superconductivity~
\cite{Micnas1,Micnas2}.
Since, in this model, the bare energy of  these electrons is dispersionless,  superconductivity must be
the result of the hybridization between these localized electrons and a wide electronic band of $c$-electrons~\cite{Werner,Hewson,Daniel1}.  Here we consider an extended or two-band Anderson type model, since we add an exchange interaction $J_{ij}$ and include a small dispersion for the $f$-electrons. For the extended model considered here hybridization  can also be detrimental to superconductivity and the competition between these two effects of the mixing is one of the important mechanisms contributing to the appearance of the SQCP.
Another important ingredient that may give rise to  superconducting quantum critical behavior~\cite{Sheila,Gegen,
Aline} in strongly correlated systems is of course the competition between repulsive and attractive interactions. Then, it is very important to take into account and discuss the effect of this additional competition and how it interferes with that associated with hybridization in giving rise to a SQCP.

We focus on different aspects of the problem, with special emphasis on the zero temperature phase diagrams and the existence and nature of any eventual SQCP. We also investigate the crossover in the lattice model from weak coupling
BCS superconductivity to Bose-Einstein condensation of pairs, as a function of the inter-site attractive
interaction.
We obtain both $s$ and $d$-wave stable SC states in different parameter regions, while most of the authors have studied only $d$-wave SC in the presence of strong local repulsive interactions~\cite{Spalek}.

It is not our purpose here to apply our results for any specific system but to identify general mechanisms that can destroy superconductivity in strongly correlated multi-band superconductors. Our study is limited to zero temperature properties and since the superconducting quantum critical points we discover are most probably above the upper critical dimension~\cite{Mu}, the slave boson mean-field approach that we use gives a reasonable description of the ground state properties.
On the other hand our approach is clearly inappropriate to describe the ground state of the cuprates in the underdoped region of the phase diagram where at finite temperatures there is evidence for a pseudogap.

\section{The model}

We consider a two-dimensional, inter-site  attractive, two-band lattice model  where local repulsive correlations are explicitly assumed between the $f$-electrons. Notice that $f$-electrons here refer generically to the quasi-particles in the narrow band. They can be either $d$-electrons as for the Fe and Cu superconductors, or $f$-electrons as for the actinides and rare-earth heavy fermions.
The Hamiltonian of the model is given by,
\begin{align}
\label{hamilt}
\mathcal{H}&=\sum\limits_{k,\sigma}\epsilon_{k}^{c}c_{k,\sigma}^{\dagger}c_{k,\sigma}+\sum\limits_{k,\sigma}
\epsilon_{k}^{f}f_{k,\sigma}^{\dagger}f_{k,\sigma}+V\sum_{i,\sigma}(c_{i,\sigma}^{\dagger}f_{i,\sigma}+h.c.)\notag\\
&\quad+U\sum_{i}f_{i,\uparrow}^{\dagger}f_{i,\uparrow}f_{i,\downarrow}^{\dagger}f_{i,\downarrow}+\frac{1}{2}\sum_{\langle i j\rangle,\sigma}J_{ij}f_{j,\sigma}^{\dagger}f_{j,-\sigma}^{\dagger}f_{i,-\sigma}f_{i,\sigma},
\end{align}
where $ c_{i\sigma}^{\dagger}$ and $ f_{i\sigma}^{\dagger}$ are creation operators for the $c$ and $f$ electrons in the wide, uncorrelated band, and in  the narrow band, respectively. These bands are described by the dispersion relations, $\epsilon_{k}^{c}$ and $\epsilon_{k}^{f}$, for $c$ and $f$-electrons in an obvious notation. The $\langle i,j \rangle$ refer to lattice sites and $\sigma$ denotes the spin of the electrons. $U$ is the on-site repulsive interaction
($U>0$) among $f$-electrons. The two types of electrons are hybridized, with a $k$-independent matrix element $V$~
\cite{Daniel1,Igor1}. This one-body mixing term can be tuned by external parameters such as pressure permitting
the exploration of the phase diagram and quantum phase transitions of the model. The last term explicitly describes
an effective attraction between $f$-electrons in neighboring sites  {\bf($J_{ij}>0$)}, which is responsible for superconductivity~\cite{Sacramento}. Notice that this term also describes  antiferromagnetic (AF), $xy$-type, exchange interactions between these electrons, such that,  magnetic and superconducting ground states are in competition. In this work we are only interested in the latter.
 We have neglected in this interaction an Ising term that when decoupled in the superconducting channel leads to $p$-wave pairing that is not considered here.
It is worth to point out that all terms included in Eq. (\ref{hamilt}) are of main importance in
influencing qualitatively the superconducting properties. The order of magnitude of these terms  can vary substantially for one specific class of systems to another.
In particular the  interactions $J_{ij}$ should be small for the case of rare-earth heavy fermions due to the localization of the $f$-orbitals in these systems. We have neglected
an inter-band attractive interaction among the $c$ and $f$-electrons,  and an intra-band term between the $c$-electrons. Also we did not include~\cite{kei} an inter-band pair hopping term which arises in second order in the hybridization when applying a Schrieffer-Wolf transformation for the Anderson lattice model~\cite{masuda}. Inter-band pairing  gives rise to ground states with finite $q$-pairing states but also to anisotropic $s$-wave superconductivity~\cite{masuda}. The main difference concerning the effect of hybridization for intra-band and inter-band pairing is that in the latter case hybridization favors superconductivity~\cite{masuda} while for intra-band pairing it is mostly deleterious.

Slave-boson formalism has been introduced to deal with strongly correlated systems under some
constraints, considering a projection method that employs \textit{slave} bosonic particles~
\cite{SB1,Coleman}.
While, in the original version, slave bosons were referring only to empty and doubly occupied states
at any given lattice site, the method was later extended by Kotliar and Ruckenstein (KR)~\cite{KR} who
also assigned slave bosons to the singly occupied states. This four-slave-boson method maps the
physical fermion (destruction) operator $f_{i,\sigma}$, with spin component $\sigma$ at site $i$,
onto the product of a (pseudo) fermion $f_{i,\sigma}$ and a bosonic operator $Z_i$. This formalism is
especially suited to deal with models of strongly correlated systems, like the ALM and the
Hubbard model, in principle,  for any value of the strength of the interactions. Besides, it
is also particularly appealing for the treatment of magnetic phases, which can already be approached
at the mean-field level owing to the presence of the single-occupancy bosons. For these
reasons, the method has been generally adopted to treat SCES, both at its mean-field level~
\cite{book1} and with the inclusion of fluctuations~\cite{Lavagna}. Specifically, it has
been found that the KR slave-boson mean-field solution is in remarkable agreement with more
elaborated Monte Carlo results over a wide range of interactions and particle densities~\cite{Lilly}.
Thus, armed with this powerful tool, we will examine the competition between superconductivity and metallic or insulating states due to the presence of hybridization and different types of interactions in multi-band systems.
Although a similar approach was carried out for the superconducting properties of the ALM~\cite{Sacramento}, here we consider a more realistic model where the correlated band has a finite dispersion.

\section{Slave bosons formalism}

Let us now come to the derivation of the effective Hamiltonian in the KR slave bosons formalism. Considering
a finite on-site interaction $U$, $f$-states can be empty, singly or doubly occupied on each site, i.e., the number of $f$-electrons per site, $n_f$ can be  $n_f=n_
{i\uparrow}^{f} +n_{i \downarrow}^{f}$=0, 1 or 2.
In order to describe all these states that the $f$-electrons can occupy, KR introduced four
bosons $e$, $d$, $p_\uparrow$, and $p_\downarrow$, where $e$, $d$ are associated with empty and doubly occupied
sites, respectively, and the bosons $p_{\uparrow}$ ($p_{\downarrow}$) with a singly occupied site with spin
component $\uparrow$ ($\downarrow$).
For the purpose of establishing a one-to-one correspondence between the original Fock space and the enlarged one, which also contains
the bosonic states, the following constraints must be satisfied:
\begin{align}\label{constraint1}
1&=e_i^{\dagger}e_i+p_{i,\uparrow}^{\dagger}p_{i,\uparrow}+p_{i,\downarrow}^{\dagger}p_{i,\downarrow}+d_i^{\dagger}
d_{i},\\
\label{constraint2}
f_{i \sigma}^{\dagger}f_{i \sigma}&=p_{i \sigma}^{\dagger}p_{i \sigma}+d_{i}^{\dagger}d_{i},
\end{align}
where  Eq.~(\ref{constraint1}) represents the completeness of the bosonic operators and Eq.~(\ref{constraint2}) the
local particle (boson+fermion) conservation at the $f$ sites. The constraints, Eq.~(\ref{constraint1}) and  Eq.~(\ref{constraint2}) are imposed in each
site by the Lagrange multipliers $\lambda_i$ and $\alpha_{i,\sigma}$, respectively. In the physical subspace, the
operators $f_{i,\sigma}$ are mapped, such that, $f_{i,\sigma} \rightarrow f_{i,\sigma}Z_{i,\sigma}$ where $Z_{i,\sigma}$ is
defined as,
\begin{equation}\label{Z}
Z_{i,\sigma}=\frac{(e_{i}^{\dagger}p_{i,-\sigma}+p_{i,\sigma}^{\dagger}d_{i})}
{\sqrt{(1-d_{i}^{\dagger}d_{i}-p_{i,\sigma}^{\dagger}p_{i,\sigma})(1-e_{i}^{\dagger}e_{i}-p_{i,-\sigma}^{\dagger}p_{i,-\sigma})}}.
 \end{equation}
The square root term in Eq. (\ref{Z}) ensures that the mapping becomes trivial at the mean-field level in the non
-interacting limit ($U\rightarrow 0$). The usual procedure consists in taking a mean-field approach
where we assume the slave bosons to be condensed~\cite{Dorin}. Then all bosons operators are replaced by their expectation values as,
$Z=\langle Z_{i,\sigma}^{\dagger}\rangle=\langle Z_{i,\sigma}\rangle=Z_{\sigma}$,
$e=\langle e_i\rangle=\langle e_i^{\dagger}\rangle$,
$p_{\sigma}=\langle p_{i,\sigma}\rangle=\langle p_{i,\sigma}^{\dagger}\rangle$, and
$d=\langle d_i\rangle=\langle d_i^{\dagger}\rangle$.
Due to translation invariance these expectation values take the same value on all sites.

Neglecting any form of magnetic order, the model given by Eq. (\ref{hamilt})
at mean-field level reduces to,
\begin{align}\label{effmodel}
\mathcal{H}_{eff}&=\sum_{k,\sigma}(\epsilon_{k}^{c}-\mu)c_{k,\sigma}^{\dagger}c_{k,\sigma}+
\sum_{k,\sigma}(\tilde{\epsilon}_{k}^{f}-\mu)f_{k,\sigma}^{\dagger}f_{k,\sigma}+\sum_{k,\sigma} Z V (c_{k,\sigma}^{\dagger}f_{k,\sigma}+h.c.)\nonumber\\&\quad
-N\frac{|\Delta|^{2}}{J}+\frac{Z^{2}}{2}\sum_{k,\sigma}(\Delta \eta_{k}f_{k,\sigma}^{\dagger}f_{-k,-\sigma}^{\dagger}+h.c.)+\lambda\sum_{k,\sigma}(p_{\sigma}^2+p_{-\sigma}^2)\nonumber\\&\quad
-\alpha\sum_{k,\sigma}(p_{\sigma}^2+d^2)+N\lambda(e^2+d^2-1)+NUd^{2},
\end{align}
where $\epsilon_{k}^{c}=-2t(\cos (k_x a)+\cos (k_y a))$ in 2D square lattice, $\epsilon_{k}^{f}=\epsilon_{0}^{f}+\gamma\epsilon_{k}^{c}$, and $\tilde{\epsilon}_{k}^{f}=\epsilon_{k}^{f}+\alpha$ is the renormalized energy level of the $f$ band,  being $\epsilon_{0}^{f}$ its bare energy level. We have considered homothetic bands, such that $\gamma=t_f/t$ ($\gamma<1$) is the ratio of the hopping terms in the $c$ and $f$-bands, and the lattice parameter $a=1$. Also in the 2D square lattice, $\eta_k=\cos k_x+\cos k_y$ and $\eta_k=\cos k_x-\cos k_y$ for $s$-wave and $d$-wave symmetries, respectively.

The $c_{k,\sigma}$ and $f_{k,\sigma}$ destruction operators refer to conduction and $f$-electrons and obey the usual anti-commutation relations, $N$ is the number of lattice sites  and $\Delta=\frac{Z^{2}J}{N}\sum_{k}\eta_{k}\langle f_{-k,-\sigma}f_{k,\sigma}\rangle$ represents the superconducting order parameter for $s$ or $d$-wave symmetry. We have added also the chemical potential $\mu$ to fix the total electronic density,  $n_{\mathrm{tot}}$. Besides, since we are interested in non-magnetic solutions, we have neglected the Hartree-Fock correction, $U\langle n_f \rangle$ to the energy of $f$-quasiparticles. To add this term simply will shift the energy level of the $f$-band, which in any case is determined self-consistently.

\subsection{Spectrum of Excitations}
\label{spectrum}
Within a BCS decoupling for the many-body attractive interaction, the quasi-particle excitations in the
superconducting phase of the model described by Eq. (\ref{hamilt}) can be obtained exactly. For this
purpose, we use the equations of motion for the Green's function~\cite{Daniel15,Daniel,Daniel15_1}. The poles of the
Green's functions yield the spectrum of excitations in the superconducting phase. The energies of these
modes are given by, $\pm\omega_{1,2}$, where,

\begin{align}\label{Bogoliubov}
\omega_{1,2}&=\sqrt{A_k\pm\sqrt{B_k}}\\
A_k&=\frac{{\varepsilon_{k}^{c}}^{2}+{\varepsilon_{k}^{f}}^{2}}{2}+\tilde{V}^{2}+\frac{(\tilde{\Delta}\eta_k)
^{2}}{2},\\
B_k&=\left(\frac{{\varepsilon_{k}^{c}}^{2}-{\varepsilon_{k}^{f}}^{2}}{2}\right)^2+\tilde{V}^{2}
(\varepsilon_{k}^{c}+\varepsilon_{k}^{f})^{2}+\frac{(\tilde{\Delta}\eta_{k})^{4}}{4}\nonumber\\
&\quad-\frac{(\tilde{\Delta}\eta_k)^{2}}{2}({\varepsilon_{k}^{c}}^{2}-{\varepsilon_{k}^{f}}^{2})+(\tilde{\Delta}\eta_k
\tilde{V})^{2},
\end{align}
where, $\tilde{V}=Z V$, $\tilde{\Delta}=Z^{2}\Delta$, $\varepsilon_{k}^{c}=\epsilon_{k}^{c}-\mu$,
and $\varepsilon_{k}^{f}=\tilde{\epsilon}_{k}^{f}-\mu$.

Following the slave boson mean-field approximation, we
replace all boson operators by their expectation values, such that, the $Z$ operator given by Eq. (\ref{Z})
becomes, $Z=p(e+d)/\sqrt{(d^{2}+p^{2})(e^{2}+p^{2})}$.
The parameters introduced, $e$, $p$, $d$, $\alpha$ and $\lambda$, can then be obtained
by minimization of the ground state energy of Eq.~\eqref{effmodel}~\cite{Sacramento}. This  together with
the number  and  gap equations,
\begin{align}
\label{nT0}
n_{\mathrm{tot}}&=1+\frac{1}{N}\sum_k \sum_{\ell=1,2}\frac{(-1)^{\ell}} {2\sqrt{B_k}}\frac{1}{2\omega_{\ell}}\Bigg \{\left(\varepsilon_{k}^{c}+\varepsilon_{k}^{f}\right)\left(\omega_{j}^{2}+\tilde{V}^{2}-\varepsilon_{k}^{c}\varepsilon_{k}^{f}\right)-
\tilde{\Delta}^{2}\eta_{k}^{2}\varepsilon_{k}^{c}\Bigg \},\\
\label{gapeqT0}
\frac{1}{J}&=\frac{Z^{4}}{N}\sum_{k}\sum_{\ell=1,2}\frac{\eta_{k}^{2}(-1)^{\ell}} {2\sqrt{B_k}}\left(\frac{\omega_{\ell}^{2}-{\varepsilon_{k}^{c}}^{2} }{2\omega_{\ell}}\right),\
\end{align}
respectively, yield a set of equations that will be solved self-consistently.

\section{Analysis of Results}

From the solution of the self-consistent coupled equation we study the zero temperature phase
diagram of the model taking into account both $s$-wave and $d$-wave symmetries of the superconducting order parameter.
In all figures below, we assume $\epsilon_{0}^{f}=0$, $\gamma=0.1$. Furthermore, we renormalize all the
physical parameters by the $c$-band hopping term  $t=1$.

\subsection{Superconducting order parameter as a function of the on-site Coulomb repulsion  and pairing interaction}
\label{repulsion}

\begin{figure}[ht]
\centering
\includegraphics[width=0.8\columnwidth]{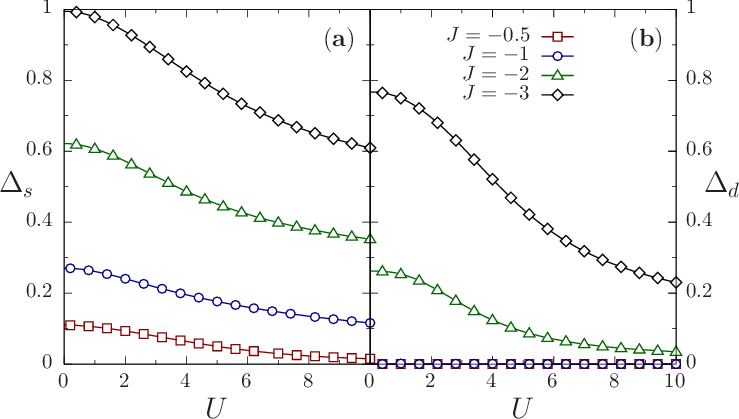}
\caption{(Color online) $\Delta_s$ and $\Delta_d$ as functions of $U$ for several values of $J$ and fixed values of
  $V=1$ and $n_{\mathrm{tot}}=2$.}
  \label{DsvsUvarJ}
\end{figure}

\begin{figure}[ht]
\centering
\includegraphics[width=0.8\columnwidth]{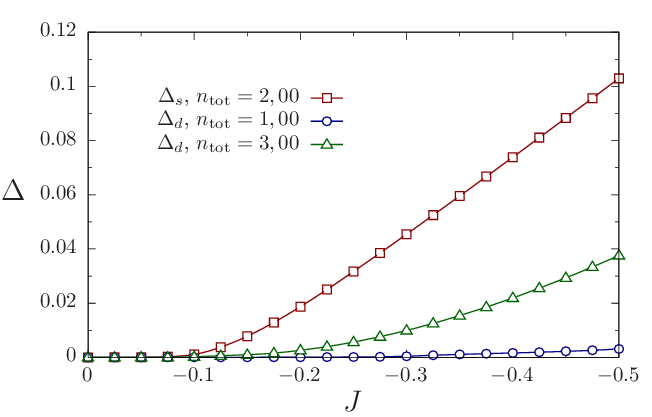}
\caption{(Color online) $\Delta_s$ and $\Delta_d$ as functions of the pairing interaction $J$ for different values of the total number of particles.We used  $U=1$, $V=1$, $\epsilon_f=0$ and $\alpha=0.1$.}
\label{DsvsJ}
\end{figure}

Figs.~\ref{DsvsUvarJ}({\bf a}) and~\ref{DsvsUvarJ}({\bf b}) show the superconducting order parameters as  functions of the repulsive
on-site Coulomb interaction $U$, considering $s$-wave and $d$-wave symmetries, respectively.
The order parameters are displayed for several values of the attractive inter-site interaction $J$ and for fixed parameters
$n_{\mathrm{tot}}=2$ and $V=1$.
Both figures show that the on-site Coulomb repulsion is detrimental for superconductivity, and specifically
in the $d$-wave case this influence is more notorious than for the $s$-wave case.
However, in both cases  we observe a monotonic decrease of the superconducting order parameter as the intensity of the local Coulomb repulsion increases, with no sign of critical behavior.  The results of Figs.~\ref{DsvsUvarJ}({\bf a}) and~\ref{DsvsUvarJ}({\bf b}) were obtained for $n_{\mathrm{tot}}=2$ at which superconductivity is rather unstable, as will be discussed below. As shown in these figures even large values of $U$ do not totally suppress superconductivity but rather cause a continuous and progressive decrease of the amplitude of the Cooper pairs. This suggests that  the suppression of superconductivity by $U$ may not be associated with a quantum critical phenomenon, at least within our mean-field  solution. In any case, $U$ is a difficult parameter to control experimentally in condensed matter physics.
A different situation will appear when we discuss the effect of hybridization on superconductivity in the section \ref{hybri}. Fig.~\ref{DsvsUvarJ}({\bf b}) seems to suggest that there is a minimum critical value of $|J|$ to support $d$-wave superconductivity, even in the absence of Coulomb repulsion. Below we show that this is not really the case.
 Fig.~\ref{DsvsJ} shows the increase of the gaps for different symmetries as functions of the pairing interaction for different total number of electrons and typical fixed values of $U$ and $V$. The gap rises smoothly as the absolute value of $|J|$ increases. We have checked that for values  of $|J| \gtrsim 0.2$ this rise is exponential with $1/J$. This is expected from the mean field character of our approximations. For smaller values of $J$ there are deviations from this behavior probably due to numerical accuracy. Notice that the same mean field approximation  implies that in case there is a critical value of $J$, that the order parameter should grow as $\Delta \propto (J-J_c)^{\beta}$, with $\beta=1/2$, which is clearly not the case. We notice from this figure that $\Delta_s$ and $\Delta_d$ can attain physical values for not too large values of $J$. The $d$-wave gap for $n_{\mathrm{tot}}=1$ is the most unfavorable,  requiring large values of $|J|$ to produce a sizable $T_{c}$ (assuming $T_{c} \propto \Delta$).

\subsection{Superconducting order parameter as a function of the total density of particles}

\begin{figure}[ht]
\includegraphics[width=0.8\columnwidth]{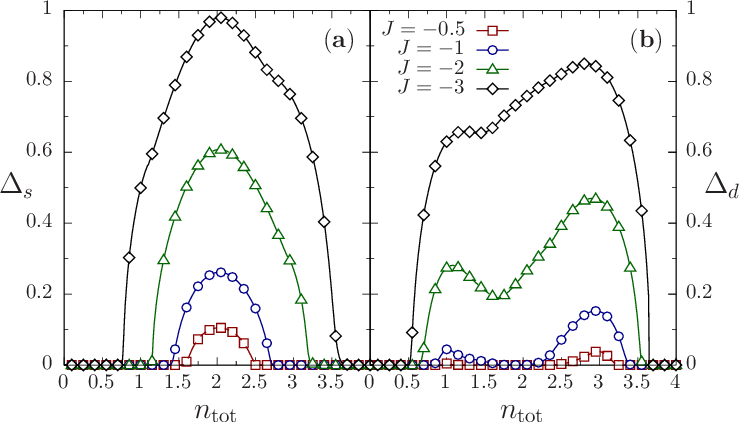}
\centering
\caption{(Color online) The zero temperature superconducting
  order parameters $\Delta_s$ and $\Delta_d$  as  functions of the band-filling $n_{\mathrm{tot}}$ for several values of $J$ and fixed values of
  $U=1$ and $V=1$.}
  \label{DsvsN}
\end{figure}

\begin{figure}[ht]
\includegraphics[width=0.8\columnwidth]{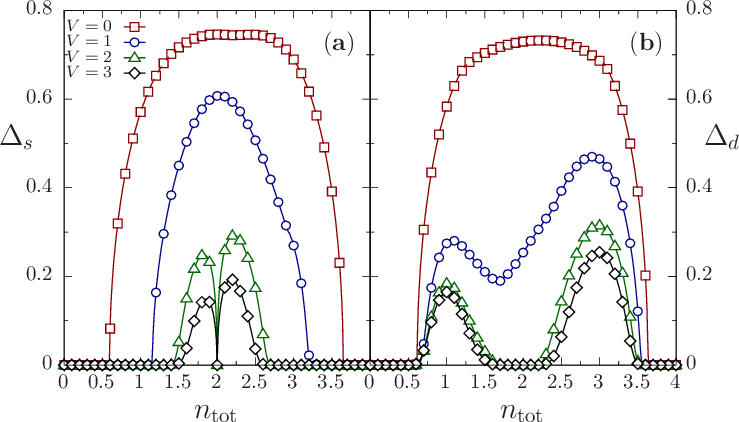}
\centering
\caption{(Color online) $\Delta_s$ and $\Delta_d$ as functions of the band-filling $n_{\mathrm{tot}}$ for several values of $V$ and fixed values of
$U=1$ and $J=-2$.}
\label{DsvsNvarV}
\end{figure}

Figs.~\ref{DsvsN}({\bf a}) and~\ref{DsvsN}({\bf b}) show the superconducting order parameters as  functions of the total density of particles
$n_{\mathrm{tot}}$, for $s$-wave and $d$-wave symmetries, respectively. They are displayed for several values of the attractive
inter-site interaction $J$ and considering $U=1$ and $V=1$.
For $s$-wave symmetry we obtain a rather symmetric superconducting dome, at least for $|J| \le 2$,  with an optimal density around $n_{\mathrm{tot}}\approx 2$. However, in the $d$-wave case
the optimal densities are close to $n_{\mathrm{tot}}\approx 1$ and $n_{\mathrm{tot}}\approx 3$.
We have calculated the contributions for the pairing amplitudes from the different electrons. We find that the $f$-electrons contribute most for the pairing, specially at the borders of the dome, i.e., for small and large $n_{\mathrm{tot}}$. This is the case for both $s$ and $d$-wave symmetries.
As expected, increasing the strength of the attractive interaction increases the region of the superconducting phase in the phase diagram. Taking the amplitude of the order parameters as a measure of the stability of the different $s$ or $d$-wave ground states, we can easily verify that $s$-wave ordering is preferred for occupations close to $n_{\mathrm{tot}}\approx 2$, while a $d$-wave type superconductivity is favoured for occupations corresponding to a total number of particles close to $n_{\mathrm{tot}}\approx 1$ and $n_{\mathrm{tot}}\approx 3$. This criterion for stability among these phases is clearly valid, at least for the weak coupling regime where the critical temperature is directly proportional to the amplitude of the order parameter. We will return to this point when we present the results for the ground state energy of the model  in Section~\ref{geenergy}.
In Figs.~\ref{DsvsNvarV}({\bf a}) and~\ref{DsvsNvarV}({\bf b}) we show the superconducting order parameters
for $s$-wave and $d$-wave symmetries, respectively, as  functions of $n_{\mathrm{tot}}$. However, differently from the previous figures, we now fix the values of the attractive interaction and Coulomb repulsion ($J=-2$ and $U=1$)  and vary the intensity of the hybridization $V$. As this increases, for both symmetries, the region of superconductivity in the phase diagram decreases showing the detrimental behavior for the superconducting ground state of large values of the mixing between the bands. For sufficiently large values, we notice that hybridization can more easily destroy superconductivity close to $n_{\mathrm{tot}}= 2$, for both $s$ and $d$-wave symmetries. In the $s$-wave case it substantially reduces the range of occupations for which the system is superconductor.

An extremely relevant question concerns the nature of the ground state that appears close to $n_{\mathrm{tot}}= 2$, as superconductivity is destroyed when hybridization increases. Is this a metallic or an insulating state?  We delay the answer to this question to further below when we study in more detail the influence of hybridization on superconductivity.

\subsection{Superconducting order parameter as a function of the hybridization}\label{hybri}

\begin{figure}[b!]
\includegraphics[width=0.8\columnwidth]{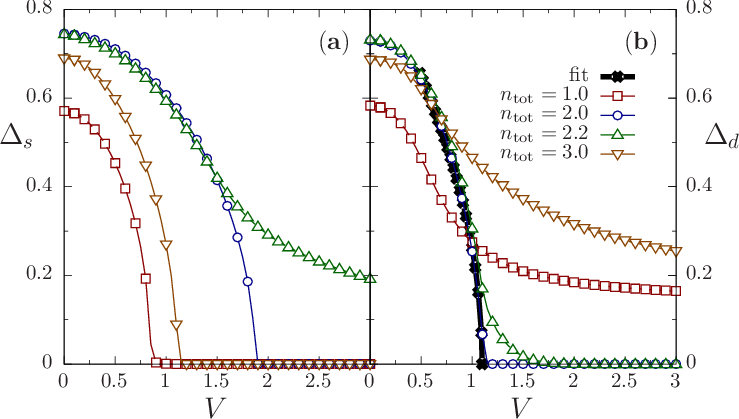}
\centering
\caption{(Color online) The zero temperature $s$-wave and $d$-wave order parameters as functions of the intensity of the
 hybridization considering several values of $n_{\mathrm{tot}}$, $J=-2$ and $U=1$. The fit in panel ({\bf b}) is to an equation, $\Delta_d \propto \sqrt{V_c-V}$, with an exponent $\beta=1/2$, as expected for the vanishing of the order parameter in a mean-field theory.}
  \label{DsvsVvarN}
\end{figure}
Figs.~\ref{DsvsVvarN}({\bf a}) and~\ref{DsvsVvarN}({\bf b}) show the zero temperature superconducting order parameters as  functions of
hybridization for several values of the filling $n_{\mathrm{\mathrm{tot}}}$, considering $U=1$ and $J=-2$
for $s$ and $d$-wave symmetries, respectively.
This phase diagram is of special interest, since hybridization
can be tuned using pressure or doping~\cite{Sheila}.

\subsubsection{\texorpdfstring{$s$}{s}-wave case}

\begin{figure}[t!]
\includegraphics[width=0.8\columnwidth]{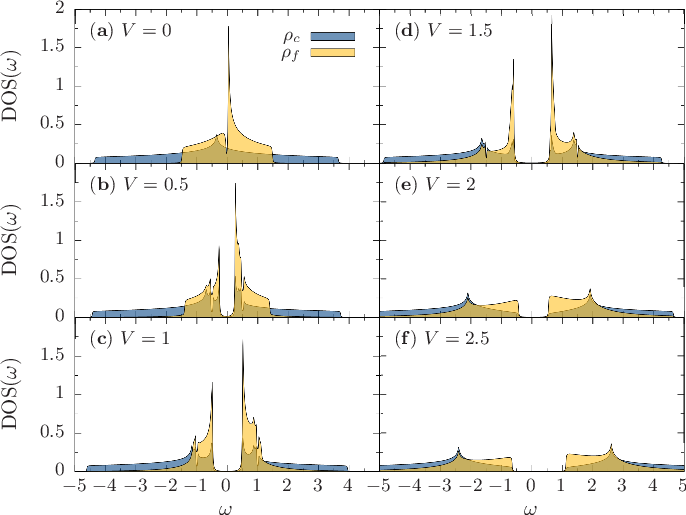}
\centering
\caption{(Color online) DOS analysis of $c$ and $f$ electrons $\rho_c(\omega)$ and $\rho_f(\omega)$ respectively,
for $s$-wave symmetry with $n_{\mathrm{tot}}=2$ and different values of $V$ for $U=1$ and $J=-2$. Notice the opening of a gap at the Fermi energy ($\omega=0$) for $V > V_c$ $(\sim1.9)$ at the   superconductor-insulator transition.}
  \label{DOSsN2}
\end{figure}

\begin{figure}[t!]
\includegraphics[width=0.8\columnwidth]{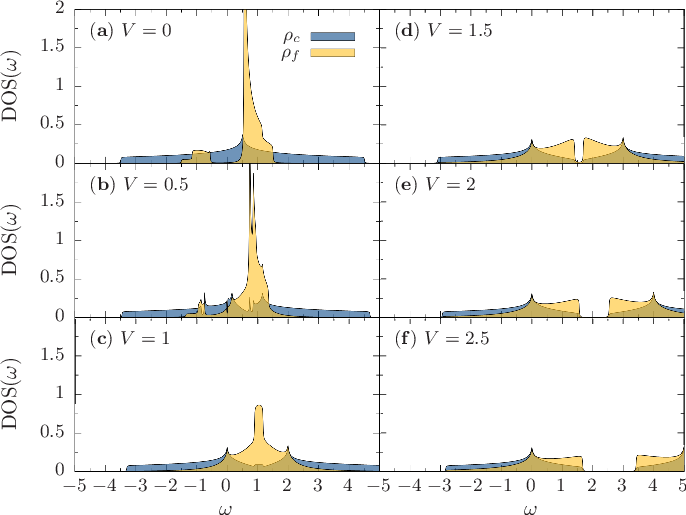}
\centering
\caption{(Color online) DOS analysis of $c$ and $f$ electrons $\rho_c(\omega)$ and $\rho_f(\omega)$ respectively,
for $s$-wave symmetry with $n_{\mathrm{tot}}=1$ and different values of $V$ for $U=1$ and $J=-2$. In this case the transition at $V_c$ $(\sim0.87)$ is from a superconductor to a strongly correlated metal as evidenced by a peak in the density of states at the Fermi level ($\omega=0$).}
  \label{DOSsN1}
\end{figure}

For the $s$-wave case, Fig.~\ref{DsvsVvarN}({\bf a}) shows clearly the suppression of superconductivity at a quantum second-order phase transition for a critical value of the hybridization $V_c$ that depends on the band-filling. The order parameter $\Delta_s \sim |V-V_c|^{\beta}$ vanishes at the SQCP at $V_c$ with a critical exponent $\beta =1/2$ as expected from the mean-field character of our approach. In order to investigate the nature of the normal state, for $V >V_c$, after superconductivity is destroyed, we calculate the density of states per spin direction (DOS) of the system in this region of the phase diagram. At half-filling, i.e., for $n_{\mathrm{tot}}=2$, the transition is to an insulating state with a gap at the Fermi level in the normal phase for $V > V_c$, as can be verified from the DOS shown in Fig.~\ref{DOSsN2}. Besides, for $n_{\mathrm{tot}}=2.2$ the superconducting phase remains finite as $V$ increases and we do not have
any evidence of superconductivity die in a SQCP. We will return to this point latter.
On the other hand, for lower and higher fillings, for instance, for $n_{\mathrm{tot}}=1$ and  $n_{\mathrm{tot}}=3$ the transition as hybridization increases is from  an $s$-wave superconductor to a metallic state. This can again be verified from an analysis of the DOS (Fig.~\ref{DOSsN1} for $n_{\mathrm{tot}}=1$). Furthermore this metal has a peak in the density of states at the Fermi level at $\omega=0$, a feature associated with correlated electronic systems.
Our results for the nature of the hybridization induced quantum superconductor-to-normal phase transition in the case of $s$-wave symmetry can be summarized by the density plot shown in Fig.~\ref{NvsVU1}({\bf a}) for $\Delta_s$ as a function of hybridization, for different band-fillings and fixed values of the Coulomb repulsion $U=1$, and the attractive interaction $J=-2$. An analysis based on calculation of the DOS as discussed above shows that, except for $n_{\mathrm{tot}}=2$, the transition to the normal state is always to a metallic state. Only when the commensurability condition of one electron per site per orbital is satisfied the normal state for large values of $V$ is an insulator with a band gap at the Fermi level. We notice from Fig.~\ref{NvsVU1}({\bf a}) that in the vicinity of this {\it half-filling} occupation there are regions of superconductivity which are quite robust, surviving for large values of the mixing. For these near half-filling occupations the decay of the order parameter with $V$ is smooth. This will be shown for the $d$-wave case discussed next, although in this case this type of smooth behavior occurs for different occupations.

\subsubsection{\texorpdfstring{$d$}{d}-wave case}

\begin{figure}[t!]
\includegraphics[width=0.8\columnwidth]{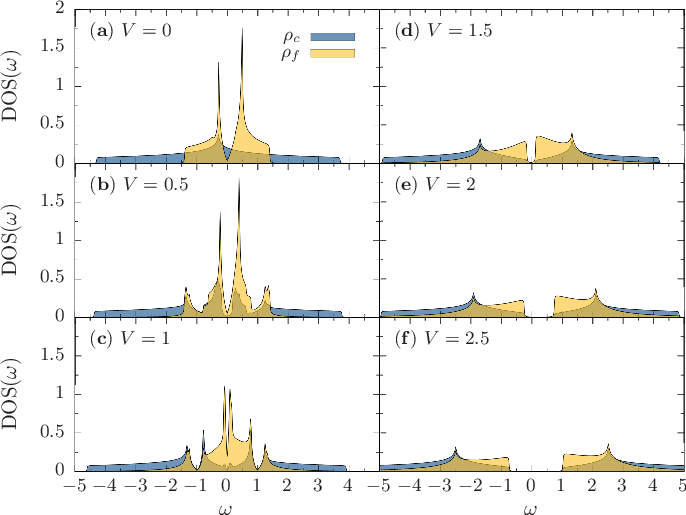}
\centering
\caption{(Color online) DOS analysis of $c$ and $f$ electrons $\rho_c(\omega)$ and $\rho_f(\omega)$ respectively,
for $d$-wave symmetry with $n_{\mathrm{tot}}=2$ and different values of $V$ for $U=1$ and $J=-2$. In this case the transition at $V_c$ $(\sim1.12)$ is a superconductor-insulator transition, with the opening of a gap at Fermi level in the normal phase.}
  \label{DOSdN2}
\end{figure}
\begin{figure}[b!]
\centering
\includegraphics[width=0.8\columnwidth]{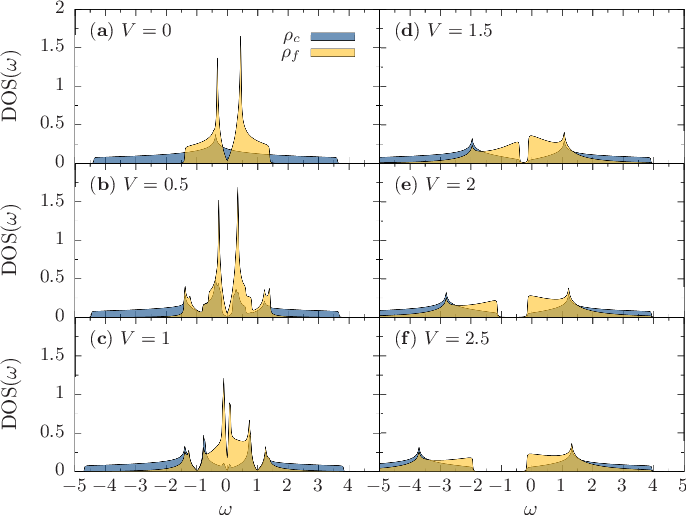}
\caption{(Color online) DOS analysis of $c$ and $f$ electrons $\rho_c(\omega)$ and $\rho_f(\omega)$ respectively,
for $d$-wave symmetry with $n_{\mathrm{tot}}=2.1$ and different values of $V$ for $U=1$ and $J=-2$. The transition at $V_c$ $(\sim1.3)$ is a superconductor-metal transition.}
  \label{DOSdN21}
\end{figure}
\begin{figure}[t!]
\includegraphics[width=0.8\columnwidth]{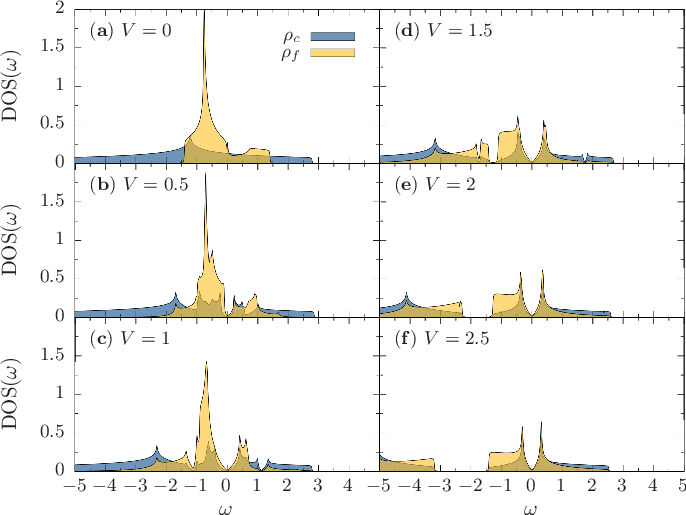}
\centering
\caption{(Color online) DOS analysis of $c$ and $f$ electrons $\rho_c(\omega)$ and $\rho_f(\omega)$ respectively,
for $d$-wave symmetry with $n_{\mathrm{tot}}=3$  and different values of $V$ for $U=1$ and $J=-2$.}
  \label{DOSdN3}
\end{figure}
\begin{figure}[t!]
\includegraphics[width=0.8\columnwidth]{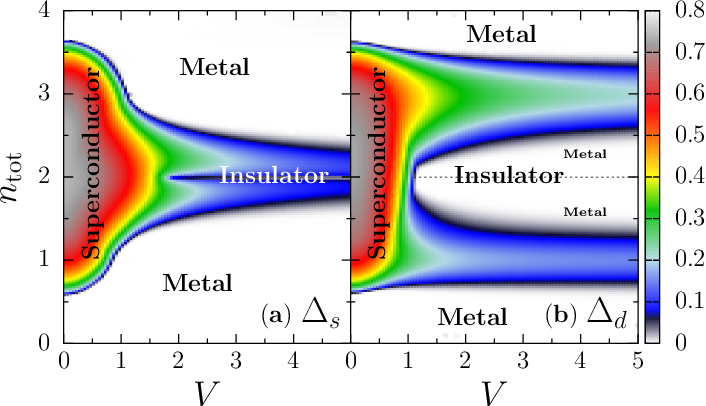}
\centering
\caption{(Color online) Density plots of $\Delta_s$ and $\Delta_d$ varying $n_{\mathrm{tot}}$ and $V$,
for fixed values of $J=-2$ and a moderate Coulomb repulsion $U=1$. The transition at $V_c$ is to an insulating state only for the commensurate filling $n_{\mathrm{tot}}=2$, for both $s$ and $d$-wave symmetries. Otherwise it is always to a metallic state. There are regions of great stability for the superconductor for $n_{\mathrm{tot}}=1$ and $n_{\mathrm{tot}}=3$ in the case of $d$-wave symmetry. For $s$-wave these regions are close to the insulator at $n_{\mathrm{tot}}\lessgtr 2$.}
  \label{NvsVU1}
\end{figure}

Fig.~\ref{DsvsVvarN}({\bf b}) shows the behavior of the zero temperature $d$-wave order parameter,  $\Delta_{d}$, as a function of $V$ for the same numerical parameters as in Fig.~\ref{DsvsVvarN}({\bf a}).
As before, from an analysis of the DOS in Fig.~\ref{DOSdN2}, we find at half-filling ($n_{\mathrm{tot}}=2$) a quantum phase transition at a critical value of hybridization, from a $d$-wave superconductor to an insulator with a gap in the density of states at the Fermi level. Besides, at any other filling
the normal state is always a metallic state. This is shown, for example for $n_{\mathrm{tot}}=2.1$, from the DOS analysis in Fig.~\ref{DOSdN21}. This quantum second-order phase transition is such that $\Delta_d \sim |V-V_c|^{\beta}$
with $\beta =1/2$ close to $V_c$.
As in the $s$-wave case we find regions where superconductivity is very stable, but now these lie close to the occupations $n_{\mathrm{tot}}=1$ and $n_{\mathrm{tot}}=3$. Along these regions, as shown in Fig.~\ref{DOSdN3} for $n_{\mathrm{tot}}=3$, the decay of the order parameter is smooth and monotonic, with no sign of a SQCP up to the large values of $V$ investigated. The same occurs for $s$-wave symmetry for band-fillings close to $n_{\mathrm{tot}}=2$.
As for the $s$-wave case, these results can be summarized by the density plot shown in Fig.~\ref{NvsVU1}({\bf b}) where the amplitude of the $d$-wave order parameter is plotted as function of the hybridization for different band-fillings.
\begin{figure}[t!]
\includegraphics[width=0.8\columnwidth]{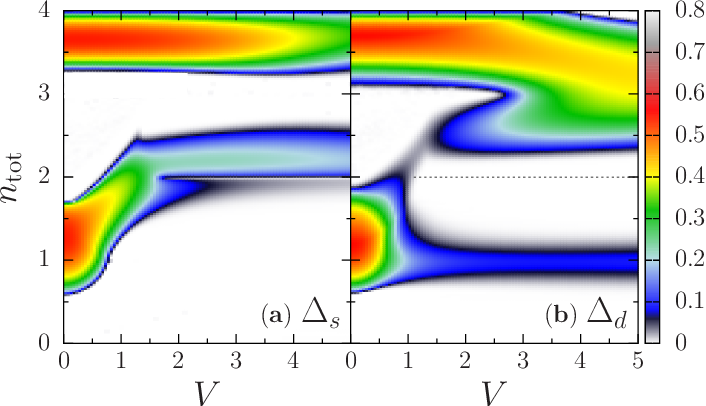}
\centering
\caption{(Color online) Density plots of $\Delta_s$ and $\Delta_d$ varying $n_{\mathrm{tot}}$ and $V$,
for fixed values of $J=-2$ and a strong Coulomb repulsion $U=10$. Notice the regions of stability associated with pairing of electrons of mainly $f$-character in doubly occupied sites for $n_{\mathrm{tot}}> 3$. }
  \label{NvsVU10}
\end{figure}
Finally, in Fig.~\ref{NvsVU10}, we show the effect of increasing Coulomb repulsion ($U=10$) on the zero temperature phase diagram of the two-band system.
When compared with the case of $U=1$, shown in Fig.~\ref{NvsVU1}, we can see substantial changes in the phase diagram. The more remarkable is the appearance of a robust region of superconductivity for band-fillings $n_{\mathrm{tot}} > 3$, for both $s$ and $d$-wave symmetries. Naively, we could expect that in this region of the phase diagram,  superconductivity would be associated with a band splitted by Coulomb repulsion from the filled singly occupied bands. However this is not the case as there is no splitted doubly occupied band in the present approach.   Superconductivity at these large values of $U$ is associated with pairing among quasi-particles in doubly occupied sites and involves essentially pairing of electrons with $f$-character.

The results above confirm and expand previous theoretical results~\cite{Daniel1,Mucio&Igor}  on the role of hybridization on superconductivity, putting in solid ground our understanding of the mechanisms for the appearance of a SQCP in multi-band correlated systems under pressure, as experimentally observed~\cite{Sheila}.

\subsection{BCS-BEC crossover}

For completeness we study the change from a Bardeen-Cooper-Schrieffer (BCS) to a Bose-Einstein condensation (BEC) type of superconductivity that occurs in our model as a function of the strength of the attractive interaction.
This may be crucial for the understanding of high $T_c$-superconductivity~\cite{Daniel15_1,Daniel15,Micnas1}, and of the pseudo-gap state~\cite{Chen}.
First we notice that in a lattice there is a natural wave-vector cutoff that avoids any possible ultraviolet divergence common to models in the continuous limit~\cite{Daniel15,Daniel15_1}. So, in this case it is not necessary to introduce any regularization procedure to study the BCS-BEC crossover

\begin{figure}[t!]
\includegraphics[width=0.8\columnwidth]{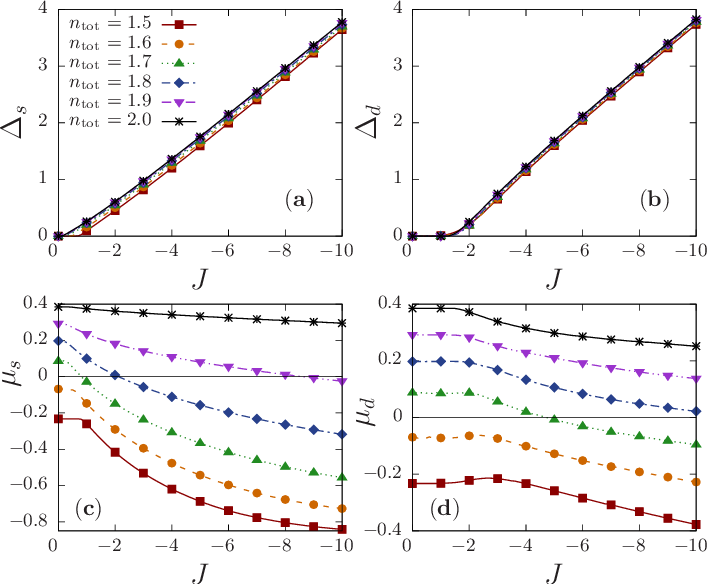}
\centering
\caption{(Color online) The zero temperature order parameters, $\Delta_s$ and $\Delta_d$, and the chemical potential for $s$ and $d$-wave symmetries as functions of the strength of the
 attractive inter-site interaction considering several values of $n_{\mathrm{tot}}$,  for $V=1$ and $U=1$.}
  \label{crossD}
\end{figure}

In Figs.~\ref{crossD}({\bf a})-({\bf d}) we plot the superconducting order parameters and
the chemical potential of the system, for several values of the filling $n_{\mathrm{tot}}$ and fixed parameters $V=1$ and $U=1$, as functions
of the intensity of the attractive inter-site interaction $J$.
In both cases the BCS-BEC crossover, characterized by an increase of the gap and decrease of the chemical potential occurs as we increase the strength of the attractive interaction. For concentrations below half-filling, we see in Fig.~\ref{crossD} that the more dilute the system the larger is the decrease of the chemical potential for negative values with increasing interaction.
Then, as it is showed in Fig.~\ref{crossD}, as $J$
increases the chemical potential drops and becomes negative signaling a  change of regime from BCS superconductivity to Bose-Einstein condensation of pairs.
It is worth to point out that the evolution from the BCS to the BEC limit
may occur also increasing the intensity of hybridization (not shown)~\cite{Daniel15}.
Notice also from Fig.~\ref{crossD}, that this crossover occurs as one dilutes the system. This can be  seen from Fig.~\ref{crossD} considering a  fixed value of the strength of the interaction and decreasing the density of quasi-particles~\cite{leggett,Duncan}.

\section{Ground state energy}
\label{geenergy}

\begin{figure}[t!]
\includegraphics[width=0.8\columnwidth]{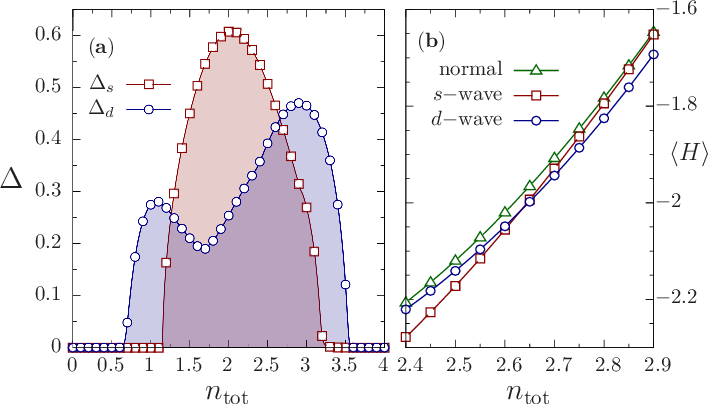}
\centering
\caption{(Color online) Analysis of the stability of the different phases that appear in the model for $V=1$ and $U=1$.
 ({\bf a}) panel shows the zero temperature superconducting order parameter, for $s$ and $d$-wave symmetries, as a function of the band-filling  $n_{\mathrm{tot}}$.
 ({\bf b}) panel: Comparing the internal energies, $\langle H\rangle$, for $2.4<n_{\mathrm{tot}}<2.9$.}
  \label{energy}
\end{figure}

We have shown above that a strongly correlated, hybridized, multi-band system may exhibit superconductivity of both $s$-wave and $d$-wave symmetries in a large region of its phase diagram. In particular,  both phases can exist in the same region of the phase diagram. A general criterion to decide the stable phase and that avoids the calculation of the ground state energy is to compare the amplitude of the order parameters associated with different symmetries. Here we check the validity of this criterion by a direct comparison with the calculated ground state energies. The results are shown in Fig.~\ref{energy}. The panel ({\bf a}) shows the amplitude of the order parameters for different symmetries as functions of the band-filling for fixed values of $U=1$ and $V=1$. We notice that according to these amplitudes, the $s$-wave symmetric superconductor is more stable for concentrations near half-filling, i.e., $n_{\mathrm{tot}}=2$. As the band-filling moves away from $n_{\mathrm{tot}}=2$ we observe an exchange of stability towards the $d$-wave superconductor. Specially for $n_{\mathrm{tot}}=1$ and $n_{\mathrm{tot}}=3$ the $d$-wave superconductor is the more stable phase. These results are directly confirmed by the calculation of the ground state energy shown in the panel ({\bf b}) of Fig.~\ref{energy}. It also shows that the superconducting phases are more stable compared with the normal phase. Then, we can conclude that the more stable, $s$ or $d$ state, corresponds to that with the larger $\Delta$ as is often the case in mean-field.

\section{Conclusions}
In this work we have studied the superconducting properties of multi-band strongly correlated systems in the presence of nearest neighbors attractive interactions. We considered that the most important interactions leading to superconductivity are those among the quasi-particles in the narrow band (intra-band case). Our model is sufficiently general to account for many systems of actual interest as high-$T_c$, Fe-pnictides and heavy fermion superconductors. Furthermore, it considers different symmetries for the order parameter. In order to treat the strong correlations we have used a generalized slave-boson approach that has proved  to be very successful for this problem. Then, we could proceed with a thorough study of our problem that had to be dealt numerically due to its complexity (seven coupled self-consistent equations). Our concern in this paper was mainly with the zero temperature phase diagram of the model as a function of the different parameters of the model. One main motivation was to investigate the existence of SQCP that can be reached and studied in real systems. This lead us to consider the influence of the strength of the mixing between different orbitals on superconductivity. For condensed matter systems this is a parameter that can be easily tuned, either by applying external pressure or doping the system. Indeed, differently from cold atom systems, $U$ and $J$ are difficult to control in condensed matter materials. We have shown that hybridization can drive a superconducting instability to an insulator, a normal metal or a correlated metal with a peak in the DOS at the Fermi level. The superconductor-insulator transition was found to occur only at half-filling, i.e., for $n_{\mathrm{tot}}=2$. For any other filling the hybridization driven superconductor-normal phase transition is always to a metallic state.
We have also investigated the effect of the local Coulomb repulsion in the superconducting state, even though this is not a realistic control parameter for condensed matter systems, as pointed out above. Its effect is deleterious to superconductivity, as expected, reducing the region for superfluidity in the phase diagram.

We have shown the relevance of the concept of BCS-BEC crossover for our study. We pointed out that it can occur by increasing the strength of the attractive interactions, but probably more relevant,  it can occur as a function of band-filling for fixed values of the attractive interaction.

We have studied the stability of the superconducting ground states with respect to the normal state and for different symmetries of the order parameters. We have shown that, in general, the $s$-wave state is preferred for occupations near half-filling, while there is an exchange of stability for a $d$-wave superconductor as the occupation goes to $n_{\mathrm{tot}}=1$ or $n_{\mathrm{tot}}=3$.

Finally, it would be very interesting to study this same problem in the presence of inter-band and pair hopping pairings since these pairing are affected differently by hybridization,
and also the Fermi surface deformation~\cite{Da2013} induced by a nodal pairing in the superconducting gap.

\section{Acknowledgments}
D.R. would like to thank the APQ1 Grant No. E-26/111.360/2014 from FAPERJ.
M.C. would like to thank the Brazilian agencies, FAPERJ and CNPq for financial support to this work. C.T. acknowledge the financial support of the Brazilian agency Capes. We would like to thank also Professor Pedro Sacramento for many enlightening discussions. The numerical calculations were performed with the aid of the cluster Ada located at the {\it Laborat\'orio de F\'isica Computacional} from IF-UFRGS.

\end{document}